\documentclass[a4paper,10pt]{article}
\usepackage{graphicx}
\usepackage{amssymb}

\begin{document}

\begin{center}

{\LARGE 
    {\bf Note on the exponent puzzle of the Anderson-Mott transition}
}

\vspace{5mm}

O. Narikiyo

{\sl Department of Physics, Kyushu University, Fukuoka 810-8560, Japan}

\vspace{3mm}

(Mar. 30, 2007) 

\vspace{5mm}

\end{center}

\begin{abstract}
The exponent puzzle 
of the Anderson-Mott transition 
is discussed on the basis of 
a duality model for strongly correlated electrons. 

\noindent
{\it Keywords}: 
{\bf Anderson-Mott transition, duality model, 
coherent-incoherent crossover}

\end{abstract}

\vspace{10mm}

The understanding of the Anderson-Mott transition 
is one of the major challendges in condensed matter physics\cite{BK}. 
Especially an exponent puzzle remains to be resolved. 

Recently the experimental determination of the critical exponent 
around the Anderson-Mott transition 
has been completed\cite{Itoh}. 
In the following 
we try to explain the crossover of the exponent 
reported in this work\cite{Itoh}. 
Such an explanation leads to the resolution of the exponent puzzle. 

We focus our attention to 
the case of nominally uncompensated systems\cite{Itoh} 
in the absence of magnetic field. 
The effect of the degree of the compensation 
will be disscussed later. 

The electrical conductivity at zero temperature $\sigma (0)$ 
around the metal-insulator transition behaves as 
\begin{equation}
\sigma (0) \propto | N/N_{\rm c} - 1 |^\mu, 
\end{equation}
where $N$ is the doping concentration 
and $N_{\rm c}$ is the critical value for the transition. 

In the case of nominally uncompensated Ge:Ga samples 
the exponent $\mu$ is experimentally evaluated\cite{Itoh} 
as $\mu \sim 1$ for $0.99N_{\rm c} < N < 1.01N_{\rm c}$ 
and $\mu \sim 0.5$ otherwise. 

To understand this exponent crossover 
we employ the duality model\cite{MN,NM,Narikiyo} 
for strongly correlated electrons. 
As a function of the energy 
the density of states for electrons is decomposed 
into two parts; coherent and incoherent components. 
The energy range for the coherent component 
is around the mobility edge $E_{\rm c}$. 
On the other hand, the energy for the incoherent component 
corresponds to that for the Hubbard band 
and is apart from $E_{\rm c}$. 
Such a model for strongly correlated electrons 
are appropriate for uncompensated case. 

To discuss the Anderson-Mott transition 
the effects of electron correlation and randomness 
should be considered at the same time. 
The relative importance of these two effects 
changes according to the degree of compensation. 
Increasing the degree of compensation 
the relative importance of randomness increases 
and that of electron correlation decreases\cite{Itoh}. 

In the case of nominally uncompensated Ge:Ga samples 
the region of $\mu \sim 1$ for $0.99N_{\rm c} < N < 1.01N_{\rm c}$ 
corresponds to the situation 
where the Fermi energy $E_{\rm F}$ is located 
at the energy for the coherent component. 
The exponent $\mu \sim 1$ is expected from the scaling theory\cite{BK} 
for the Anderson transition. 
Thus the nature of the transition is not correlation-driven Mott-type 
but randomness-driven Anderson-type and this region is understood 
as the critical region. 
On the other hand, 
$E_{\rm F}$ is located 
at the energy for the incoherent component 
in the region of $\mu \sim 0.5$. 
The exponent $\mu \sim 0.5$ results from 
the scattering with incoherent character\cite{Castner}. 
Thus this region is understood as a non-critical one 
where the incoherent component dominates. 

The width of the critical region is 
controlled by the degree of compensation\cite{Itoh}. 
In the critical region 
the effect of randomness dominates over that of correlation. 
On the other hand, 
in the non-critical region 
the correlation dominates over randomness. 

The remaining puzzle is the lack of consistency 
among the exponents\cite{Itoh} 
in the region of $\mu \sim 0.5$, 
while in the region of $\mu \sim 1$ 
the exponents are consistently understood 
by the scaling theory for the Anderson transition. 
However, our resolution of the puzzle is rather trivial. 
In the region of $\mu \sim 0.5$ 
the incoherent component dominates 
and the situation is nothing to do with the critical phenomena. 
Thus, for example, we do not worry about the relation 
between the conductivity exponent $\mu$ and 
the localization-length exponent $\nu$ 
defined by 
\begin{equation}
\xi \propto ( 1 - N/N_{\rm c} )^{-\nu}. 
\end{equation}
These two exponents need not to be related 
in the non-critical region. 
Moreover, the experimentally observed value\cite{Itoh} 
of $\nu \sim 1/3$ is easily derived by a simple argument\cite{comment}. 

In summary 
the metal-insulator transition for doped semiconductors 
is randomness-driven Anderson-type 
and the exponents for the outside of the critical region 
has nothing to do with the critical phenomena.


\begin{thebibliography}{9}
\bibitem{comment}
The localization length is identical to the cluster size 
of the localized magnetic moments. 
Such clusters are divided by mobile carriers 
whose motion is incoherent hopping\cite{Itoh} 
and whose number density is $N_{\rm c} - N$. 
Thus the localization length 
is represented by the mean distance of mobile carriers 
and proportional to $(N_{\rm c} - N)^{-1/3}$ in three dimensions. 
The same argument is employed to understand 
the incoherent nature of high-temperature superconductors 
near the Mott transition 
in Appendix of Ref. [5].
\end{thebibliography}
\end{document}